\RequirePackage{fix-cm}
\documentclass[smallextended]{svjour3}       
\smartqed  
\usepackage{graphicx}
\usepackage{array}
\usepackage[round]{natbib}

\usepackage{algorithmic}
\usepackage{algorithm}
\usepackage{url}
\usepackage{tikz}
\usepackage{soul}
\usepackage{xcolor}

\usepackage[utf8]{inputenc} 
\usepackage[T1]{fontenc} 

\usepackage{booktabs} 
\usepackage{colortbl} 
\usepackage{multirow} 
\usepackage{rotating} 
\usepackage{listings} 
\newcolumntype{L}[1]{>{\raggedright\let\newline\\\arraybackslash\hspace{0pt}}m{#1}}
\newcolumntype{C}[1]{>{\centering\let\newline\\\arraybackslash\hspace{0pt}}m{#1}}
\newcolumntype{R}[1]{>{\raggedleft\let\newline\\\arraybackslash\hspace{0pt}}m{#1}}

\definecolor{mygray}{gray}{0.85}
\definecolor{mygray2}{gray}{0.95}

\newcommand{\keyword}[1]{\textbf{#1}}

\newcommand{\code}[1]{\texttt{#1}}

\newcommand{\ie}{\emph{i.e.}, } 
\newcommand{\eg}{\emph{e.g.}, } 
\newcommand{\etal}{\emph{et al.} } 
\newcommand{\dottedname}[1]{\url{#1}}

\newcommand{\supers}[1]{\kern .01em\textsuperscript{#1}}

\newcommand{\cfr}{\emph{cfr.}}
\newcommand{\chizpurfle}{\emph{Chizpurfle}}
\newcommand{\mytilde}{{\raise.17ex\hbox{$\scriptstyle\sim$}}}

\usepackage[framemethod=tikz]{mdframed}
\mdfdefinestyle{bwstyle}{%
roundcorner = 0.5pt,%
outerlinewidth=1pt, outerlinecolor=black,%
backgroundcolor=black!7,%
align=center
}
\newcommand{\researchquestion}[2]{\begin{mdframed}[style=bwstyle]
RQ#1: #2
\end{mdframed}
}

\usepackage{marginnote}

\makeatletter

\newcommand{\rev}[1]{%
}
\makeatother

\begin{document}

\title{Evolutionary Fuzzing of Android OS Vendor System Services}


\author{Domenico Cotroneo         \and
        Antonio Ken Iannillo	  \and
        Roberto Natella
}


\institute{Domenico Cotroneo, Antonio Ken Iannillo, Roberto Natella \at
              Università degli Studi di Napoli	Federico II\\
              via Claudio 21, 80125, Napoli, Italy \\
              \email{cotroneo@unina.it}\\
              \email{antonioken.iannillo@unina.it}\\
              \email{roberto.natella@unina.it}\\\\
		   Antonio Ken Iannillo \at
		   SnT, University of Luxembourg \\
		   29 avenue J.F. Kennedy, 1855, Luxembourg, Luxembourg\\
		   \email{antonioken.iannillo@uni.lu}
}

\journalname{Empirical Software Engineering}

\date{}

\maketitle

\begin{abstract}
Android devices are shipped in several flavors by more than 100 manufacturer partners, which extend the Android ``vanilla'' OS with new system services, and modify the existing ones. These proprietary extensions expose Android devices to reliability and security issues. In this paper, we propose a coverage-guided fuzzing platform (\chizpurfle) based on evolutionary algorithms to test proprietary Android system services. A key feature of this platform is the ability to profile coverage on the actual, unmodified Android device, by taking advantage of dynamic binary re-writing techniques. We applied this solution on three high-end commercial Android smartphones. The results confirmed that evolutionary fuzzing is able to test Android OS system services more efficiently than blind fuzzing. Furthermore, we evaluate the impact of different choices for the fitness function and selection algorithm.
\end{abstract}

\section{Introduction}
\label{sec:introduction}

Android is the most common operating system for mobile devices, such as smartphones and tablets \citep{statista_2017q2}. Android is shipped in several flavors implemented by a large number of vendors, \eg Samsung, Huawei, LG, Motorola, and more than 100 other companies \citep{android_partners}. These vendors extend the Android Open Source Project (AOSP) \citep{aosp} with new and customized software services, in order to differentiate from the competition. However, these customizations also lead to reliability and security risks for the users, since they are not as well tested as the open-source version of Android and introduce additional attack vectors \citep{xu2016toward}.

In this paper, we present an approach based on fuzzing and on evolutionary algorithms to test proprietary Android system services. On the one hand, fuzzing is a well-established and effective robustness and security testing technique, that identifies weaknesses in fragile software interfaces by injecting invalid and unexpected inputs \citep{miller1990empirical}. Fuzzing was initially conceived as a black-box technique; subsequent studies showed that its effectiveness can significantly benefit from information about the test coverage, by steering the generation of test inputs towards uncovered paths \citep{bohme2016coverage,afl2016,bounimova2013billions,ossfuzz}. On the other hand, evolutionary algorithms are a popular approach to address search-based software engineering problems \citep{whitley2001overview,harman2012search}, and represent a promising solution for coverage-guided fuzzing: they can leverage coverage information to generate new inputs by evolving the most promising current ones, such as recent inputs that discovered new code paths. Since the effectiveness of tests is statistically correlated to code coverage \citep{kochhar2015code}, the measurable goal of evolutionary testing is to reach the highest coverage possible.

Applying this fuzz testing strategy on Android customizations is technically challenging, because it requires test coverage information from proprietary code. Instrumenting at compile-time the Android OS is inconvenient for Android vendors (due to the complexity of the Android build process), and impossible for independent testers (as they do not have access to the source code). Furthermore, it is also not possible to get coverage information by running the customized Android OS in an emulated environment, since customizations can often only run on the real hardware. \citep{yaghmour2013embedded} 
The lack of source code also introduces other technical challenges for test input generation, including the unfeasibility of accurately measuring the code size of the Android service under test (as several services run in the context of the same OS process, without a clear separation of the code of individual services), the ``saturation'' of the coverage of specific methods of the service under test (since, for the previous reason, coverage cannot be measured in relative terms with respect to the total code size), and the identification of the input surface of the service.

We overcome these limitations by proposing the \chizpurfle{} platform for evolutionary fuzzing of Android OS services. The contributions of this paper are the following.

\vspace{2pt}
\noindent
\textbf{A new fuzzing approach for the Android OS based on genetic algorithms}. 
We apply evolutionary computing to fuzz testing, by proposing a genetic algorithm tailored for fuzzing Android OS services. Genetic algorithms are a special class of evolutionary algorithms, where the tentative solutions (which, in our context, are represented by a set of input parameters to invoke a method of an Android OS service) are evolved by recombination and mutation \citep{back1996evolutionary}. 
However, simply applying genetic algorithms to generate test inputs can be inefficient, since there is a risk of wasting testing time on service methods that are already well-covered (but the saturation of the relative coverage cannot be accurately measured, due to the lack of source code). Therefore, we introduce the idea of \keyword{community} for co-evolving test inputs across the several, heterogeneous methods of a service under test. This approach varies the relative size of input populations within the community, and awards the most promising populations in order to carefully focus the test time budget on specific methods.

\vspace{2pt}
\noindent
\textbf{A reusable, extensible platform, \chizpurfle{}, that implements the proposed fuzzing approach}.
The \chizpurfle{} platform aims to be a basis for research on evolutionary fuzzing in the context of mobile devices. It overcomes the technical issues for applying fuzz testing strategies, including dynamic binary rewriting techniques to collect coverage information from real commercial devices and to perform coverage-guided tests on proprietary customizations. 
Moreover, the design of the \chizpurfle{} platform provides representations for the elements of genetic algorithms (individual, fitness, community, population, ...), in order to be extensible and to enable experimentation (e.g., with new heuristics, new mutation operators, ...) to further enhance evolutionary fuzzing. 
We release \chizpurfle{} as open-source software\footnote{The source code and documentation of the \chizpurfle{} platform is available at \url{https://github.com/dessertlab/fantastic_beasts}}, to allow users to experiment with their own algorithms. Since the design space of evolutionary algorithms is wide \citep{whitley2001overview,harman2012search}, we believe that such a platform can be a valuable asset for supporting future research.

\vspace{2pt}
\noindent
\textbf{An experimental evaluation of evolutionary fuzzing on several commercial Android devices, under different configurations of genetic algorithms}.
This paper presents an experimental study on three high-end commercial Android smartphones. We applied evolutionary fuzzing in 420 fuzzing campaigns over 15 different proprietary services (5 for each device), for a total of more than 8 000 000 tests, performed on the actual devices, which lasted more than 60 days (CPU time). 
Overall, evolutionary fuzzing is able to test Android services more thoroughly than black-box fuzzing: while some services may not benefit from evolutionary fuzzing (for example, services with a simpler input surface, or services that are influenced by the hardware state), for other services the improvement can be very high (the coverage doubles in several cases, and it is even higher than 24x in one specific case).  
Moreover, we expect that the effectiveness of evolutionary fuzzing can be further improved by introducing new mutation operators and new heuristics.
We also compared several configurations of evolutionary fuzzing, by considering different \emph{fitness functions} (which evaluate the quality of test inputs) and \emph{selection algorithms} (which choose the inputs to evolve). The experimental results show that some of the considered fitness functions perform better than others in sporadic cases, but they are equally effective from a statistical point of view; and that the non-parametric selection algorithms exhibit statistically-significant better performance than the parametric one.

\vspace{2pt}

Compared to our earlier work \citep{iannillo2017chizpurfle}, which adopted a simpler evolutionary approach, and which separately tested the methods of Android OS services, this paper presents a new approach based on genetic algorithms, on the co-evolution of test inputs through the idea of community, and on a new extensible architecture of the \chizpurfle{} platform. Moreover, this paper presents more extensive experimentation on several Android devices, and investigates the impact of different configurations of the genetic algorithm.

The rest of the paper is structured as follows: in Section~\ref{sec:rq}, we describe in more detail the problem of testing Android system services, and the research questions addressed in this paper; Section~\ref{sec:evo_fuzzing} presents the \chizpurfle~platform; Section~\ref{sec:results} shows the experiments on commercial Android devices; Sections~\ref{sec:threats} and \ref{sec:related_work} discuss the threats to validity and related work;  Section~\ref{sec:conclusion} concludes the paper.

\section{Problem statement and research questions}
\label{sec:rq}

The Android OS is a software stack based on a service-oriented architecture. Mobile apps consume services through the rich, high-level Android framework APIs. In turn, the classes of the Android framework invoke the \emph{system services} of the Android OS, which run in separate, privileged processes of the Android OS, and which perform the actual work (e.g., accessing hardware resources, managing the app lifecycle, interacting with other apps and the user, etc.). 
These system services provide a public interface represented by \emph{service methods}, which include a set of method parameters in input and output. This public interface is invoked by the Android framework through remote procedure calls (RPC) using the Android Binder API, in a similar way to distributed middleware. 
Examples of system services from the vanilla Android OS are: the \emph{Activity Manager}, which exposes services for managing the lifecycle of apps; the \emph{Connectivity Manager}, which abstracts the networking capabilities of the device; and the \emph{Package Manager}, which exposes services for handling and checking the permissions of apps.

\begin{figure*}
  \centering
  \includegraphics[width=\textwidth]{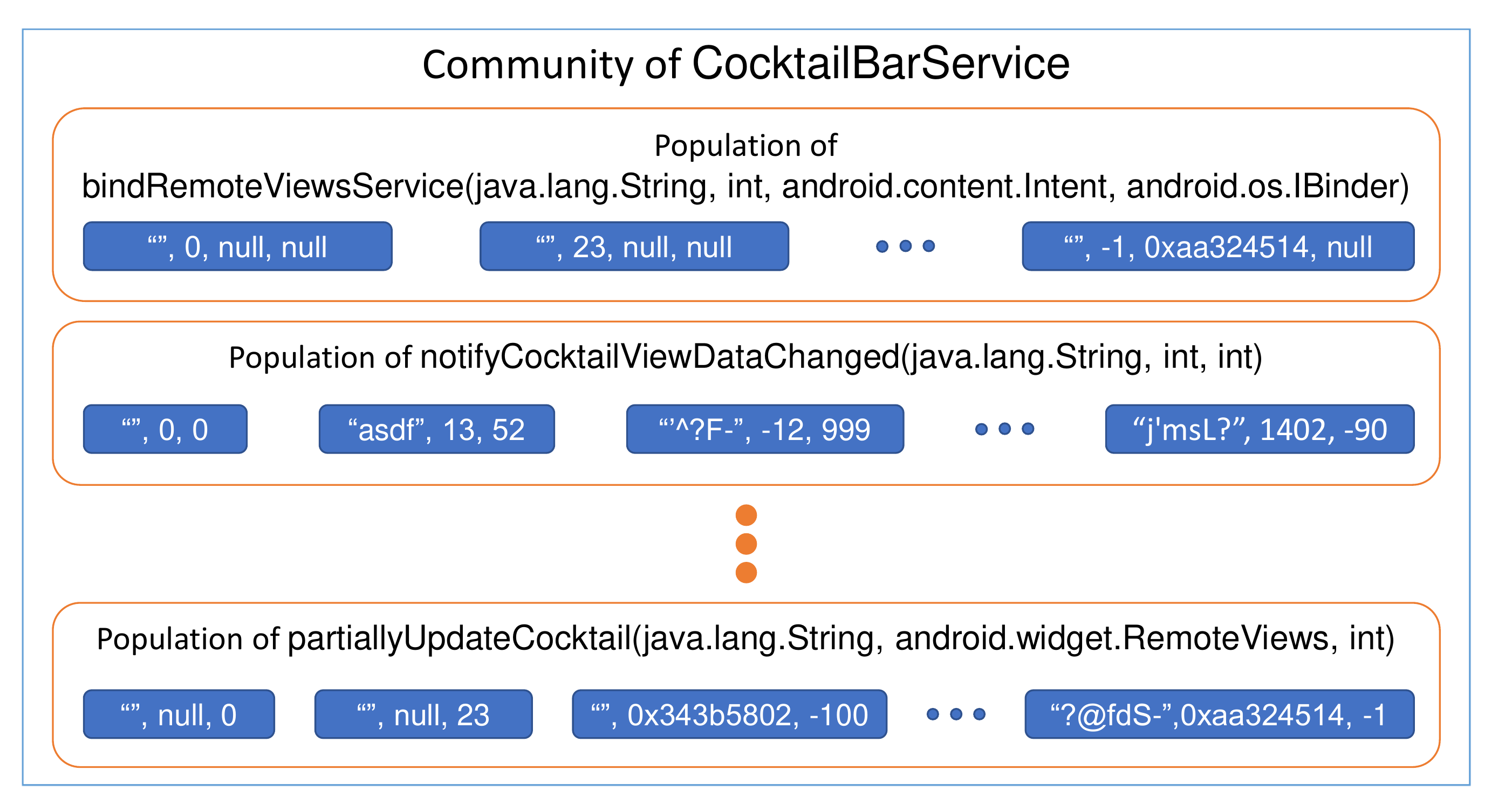}
  \caption{Individuals, populations, and community for the \code{CocktailBarService} of the Samsung Galaxy S6 Edge.}
  \label{fig:visual_example}
\end{figure*}

Android vendors modify these system services to provide support for the specific hardware of their products. For example, the \emph{Camera Service} and the \emph{Radio Interface Layer} of the vanilla Android OS are customized by linking them to custom drivers for the camera and the baseband processor. Moreover, Android vendors can add system services from scratch in order to provide new features. For example, the \code{CocktailBarService} from Samsung Galaxy S6 Edge provides APIs for handling a new type of GUI notifications, with 41 public methods (\figurename{}~\ref{fig:visual_example}).

 In order to generate effective tests for these system services, we adopt fuzzing and evolutionary algorithms.
Our fuzzing approach generates method calls with inputs that match the signature of the target method, thus, the inputs can be considered ``valid'' from the syntactic point of view (i.e., the type system of the Android service interface). However, these inputs can be ``invalid'' from the point of view of the target Android service, depending on the specific meaning of the parameters and of the method call. For example, a very large value submitted through an integer input parameter may represent an out-of-norm value for the service, or a ``null'' object reference may violate the assumption that a given input object reference should always be a valid one. 
Testing against such ``invalid'' inputs is the intended goal of fuzzing, which aims to abuse the service interface to find inputs that are not gracefully handled. The fuzzing approach generates invalid inputs by using mutation operators that introduce boundary integer values, unusual characters, null references, and combinations of thereof (e.g., by using cross-over on the inputs of two method calls). It is important to identify these invalid inputs since they pose a reliability/security threat, as they can originate from any buggy or malicious application.
Random fuzzing may be inefficient at exploring the space of service API inputs, which is typically very large; therefore, evolutionary fuzzing tunes the generation of new inputs according to the test coverage achieved by previous inputs.

In the context of evolutionary fuzzing, an \keyword{individual} (i.e., a candidate solution) is represented by a vector of input values for one of the methods of the service under test. A \keyword{population} (i.e., a group of individuals at a given round of an evolutionary algorithm) includes a set of several input vectors. A \keyword{fitness function} assigns a numerical value (\ie the fitness) to an individual, which in our context represents the potential of the individual to generate good new individuals (through mutation or cross-over) that are able to reach not-yet-executed code. Evolutionary fuzzing evolves the population over several iterations through the following steps:

\begin{enumerate}
\item a \keyword{selection algorithm} to select which individuals from the previous population to evolve;
\item a \keyword{cross-over} phase, that slices and recombines two individuals from the previous population to generate a new one;
\item a \keyword{mutation} phase, which modifies an element of an individual to generate a new version of that individual.
\end{enumerate}

In the cross-over phase, we need to avoid combining two input vectors that refer to different methods, since the resulting vector may not comply with the number and type of parameters of the method signature. Therefore, we introduce the notion of \keyword{community}. We restrict a population to include individuals for the same method of the service under test, and we include several populations in a community, which is iteratively evolved as a whole. 

In the previous example of the \code{CocktailBarService} from Samsung Galaxy S6 Edge (\figurename{}~\ref{fig:visual_example}), the community consists of 41 different populations, one for each method. In a population, every individual represents an input vector for the method. For example, the method \code{notifyCocktailViewDataChanged} has three parameters, \ie <\code{java.lang.String}, \code{int}, \code{int}>, a potential individuals in its population are <\code{"asdf"}, \code{13},\code{52}>, <\code{"j'msL?"}, \code{1402},\code{-90}>, and so on.

This organization enables the co-evolution of individuals across the service, by varying the relative size of populations within the community, and by awarding the most promising populations. This approach can make better use of the test budget and explore the target service more efficiently. For example, if a service has a method that can only reach a small part of the target service (\eg a get/set method that only reads/writes a variable inside the service), the community will eventually reward individuals for the other methods, thus steering the testing budget on the methods that have a higher potential to cover more code of the target service.

The problem of co-evolving individuals becomes even more important in the context of Android OS services. The first reason is that it is unfeasible to detect the \emph{saturation} of the tests for a method by just looking at the relative coverage of the tests, due to the lack of source code (this technical issue is further discussed in Section~\ref{sec:results}). 
The second reason is that, compared to previous evolutionary testing studies (which are mostly applied for unit testing), the Android services are complex subsystems, with several tens of methods that share a significant part of the code of the service \citep{yaghmour2013embedded,levin2015android,androidxref_aidl}.
For example, a service for SMSs can provide methods to send messages with different flavors (e.g., SMSs with simple text, multipart text, premium SMSs, etc.), which likely reuse common routines to handle low-level SMS communication. Moreover, services typically expose methods to get/set internal parameters, to register event listeners, etc. that tend to use shared data.  
Therefore, the tests for different methods will overlap at covering the shared code, and it becomes useful to co-evolve the inputs of different methods in order to drive exploration towards the non-shared code.

It is also important to consider that testing proprietary Android services must address a number of technical challenges that are different from other areas of software testing. 
One of the issues is to identify the set of customized services in the Android device, and what is the \emph{input interface} of the Android services under test (i.e., the methods of the services and their parameters), which is a basic requirement for performing any kind of testing activity. Then, before testing, we need to point out the services that were customized by the vendor of the Android device with respect to the open-source Android (AOSP). 
Another key issue is tracking the coverage of the tests even if the source code is not available, which we address by means of dynamic binary instrumentation. In this work, we provide a reusable platform that overcomes these technical issues, and that can serve the research community at focusing on better fuzz test generation strategies.

In the rest of the paper, we discuss more in detail how these concepts are implemented in our \chizpurfle{}~platform. Moreover, in the experimental part of the work, we consider the following research questions on evolutionary fuzzing.

\researchquestion{1}{Can evolutionary fuzzing improve the test coverage of Android OS system services, compared to black-box fuzzing?}

First, we want to determine whether fuzzing with genetic algorithms can repay the additional complexity and computational overhead, compared to the simpler, black-box fuzzing approach (\ie mutating input values without exploiting any coverage information). We cannot take for granted that this approach performs better than its black-box counterpart, since the performance of evolutionary algorithms is sensitive to the difficulty of the problem at hand \citep{naudts2000comparison,mitchell1992royal}. 
Therefore, the performance evaluation will compare evolutionary and black-box fuzzing with respect to test coverage, which is a key performance indicator for fuzzing techniques \citep{klees2018evaluating}. In order to assess the statistical significance of the differences, we base the comparison on hypothesis testing \citep{arcuri2014hitchhiker}. If evolutionary fuzzing can consistently deliver a better test coverage than black-box fuzzing, then it should be preferred since it will bring a higher potential to find more bugs and to increase the overall reliability of the Android device. 
Moreover, the evaluation will also analyze any unique bug found by evolutionary fuzzing.

We must consider that the tuning of parameters has a great influence on the performance of genetic algorithms \citep{grefenstette1986optimization, whitley2001overview,arcuri2014hitchhiker,fraser20151600}. 
In particular, when designing a genetic algorithm, a tester should consider two critical aspects at first, that is, the fitness function for evaluating the quality of the current solution, and the selection algorithm to selects inputs that should be prioritized for generating new ones.

\researchquestion{2}{What is the impact of the fitness function and of the selection algorithm on evolutionary fuzzing of Android OS system services?}

In the context of fuzz testing, several design choices are possible for the fitness function and the selection algorithm \citep{whitley2001overview,bohme2016coverage,afl2016,back1996evolutionary,back1991extended,goldberg1991comparative}. In the paper, we evaluate different configurations and investigate their impact on the code coverage achieved by evolutionary fuzzing.

\section{Evolutionary fuzzing approach}
\label{sec:evo_fuzzing}

We present the basic components of \chizpurfle{} in \S\ref{subsec:chizpurfle_platform}, the extended architecture of the \chizpurfle{} platform in \S\ref{subsec:extended_arch}, and the evolutionary fuzzing algorithm in \S\ref{subsec:evo_fuzz_alg}.

\subsection{The basic Chizpurfle components}
\label{subsec:chizpurfle_platform}

\begin{figure*}
  \centering
  \includegraphics[width=\textwidth]{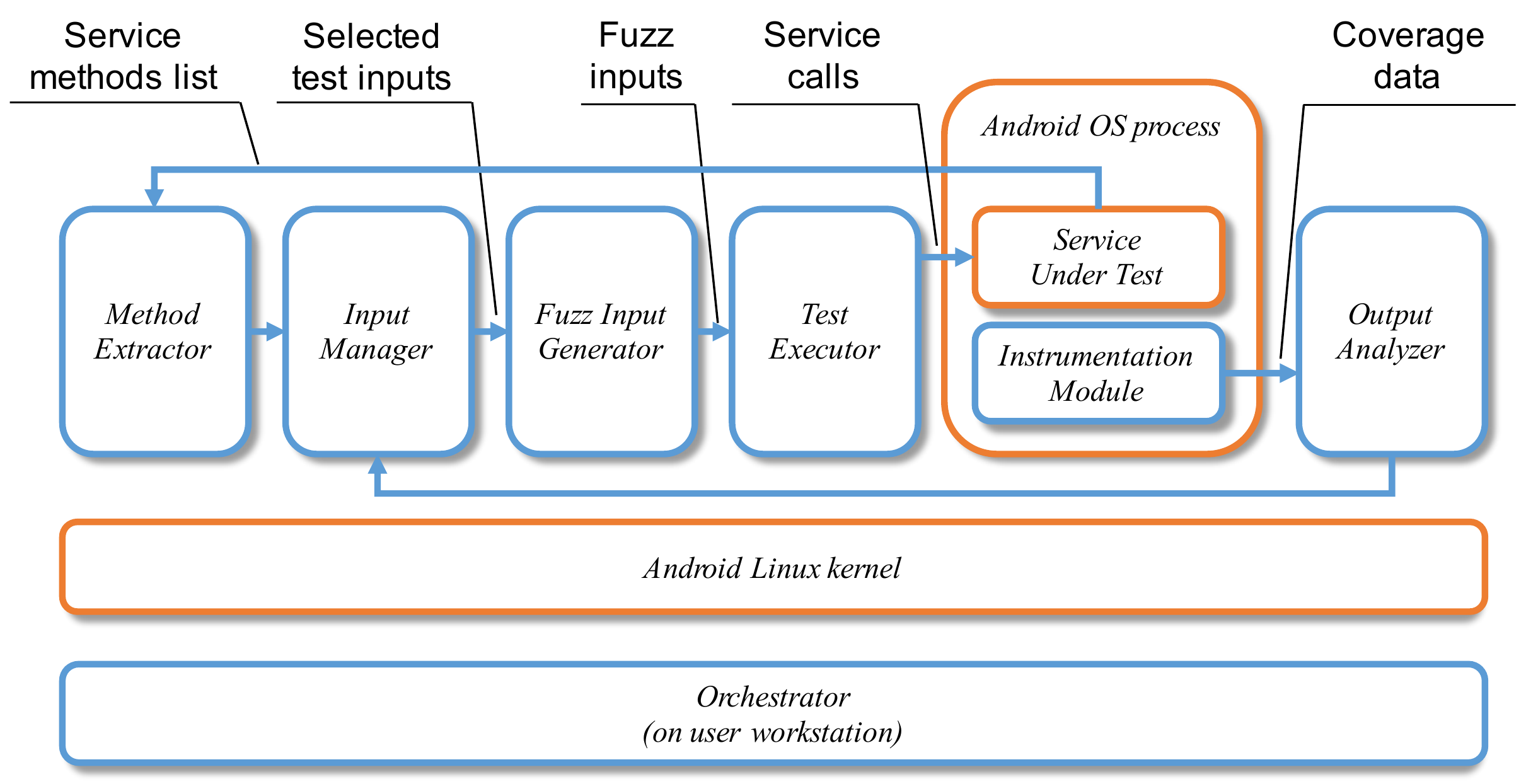}
  \caption{Overview of Chizpurfle components.}
  \label{fig:evo_platform}
\end{figure*}

\chizpurfle{} includes six software modules that run on the target Android device, that cooperate to generate fuzz inputs, to execute tests, and to profile the target system service; and an external orchestrator on the user workstation (\figurename{}~\ref{fig:evo_platform}). 
These modules were developed in the preliminary tool that we presented in previous work \citep{iannillo2017chizpurfle}, and are the basis for the evolutionary fuzzing platform presented in this work.

The \textbf{Orchestrator} controls the other modules through the Android Debug Bridge (ADB) \citep{adb}, and prevents the early termination of fuzz testing in the case of crashes of system processes.

The \textbf{Method Extractor} generates a list of services from the target Android device with vendor customizations, and it compares them with the AOSP of the same version. 
The Method Extractor queries the \emph{Service Manager} of the Android OS to get the list of all registered services, including customized ones. By iterating on these names, it retrieves the list of service descriptors. In the case of Java-implemented services (supported by the current version of the tool), a service descriptor is the string name of the Java Interface that is implemented by that system service (\emph{e.g.,} the \emph{Package Manager} service implements the \emph{android.content.pm.IPackageManager} Java Interface). Then, we use the Java Reflection API to inspect the definition of the interfaces, and to get the signatures of the methods in the service. Since the Java Reflection API is also used for managing communication in Binder, it can reliably get the list of invokable methods of the service. The methods that cannot be found in the AOSP, or that have a different signature than their AOSP counterpart, are marked as ``vendor customizations'' and considered for testing. Furthermore, it identifies the system process that hosts the target service: it forces the system services to be re-published again by restarting the system processes, and intercepts registration calls from the services to the Service Manager using the \emph{ptrace} system call of the Linux kernel, in a similar way to debugging tools, in order to identify in which process the services are running.

The \textbf{Instrumentation Module} profiles the coverage of the system service under test. We designed this component to run on the actual, unmodified Android device, and to avoid to interfere with the original behavior of the service. 
The component is attached to the Android OS process where the proprietary service is running, without neither restarting the process (as most of these services are already running since the boot of the device) nor recompiling the source code of the service.

The Instrumentation Module uses the \emph{ptrace} system call of the Linux kernel, which allows to write on the memory address space and CPU registers of a process. 
We leverage \emph{ptrace} to perform \emph{dynamic binary rewriting} of program code \citep{nethercote2007valgrind,luk2005pin,frida2017}, in a similar way to virtual machine interpreters. The program is divided in \emph{basic blocks}, which are small groups of sequential machine instructions that end with a branch. This branch instruction is replaced with a branch to the Instrumentation Module: when the branch is reached, the control flow is returned to our module, which retrieves the next basic block, applies some transformations (such as just-in-time compilation and instrumenting the final branch instruction) and moves the control flow to the block. Moreover, this process is accelerated by caching basic blocks that have been already processed, so that the exit branch can directly jump to the next basic block if it is cached. In our context, we apply this technique to keep track of which basic blocks are executed, in order to compute the test coverage.

The \textbf{Output Analyzer} gathers coverage data from the Instrumentation Module through a socket connection, analyzes and saves the data, and feeds back information to the \textbf{Input Manager}, is in charge of ranking test inputs. In turn, the \textbf{Fuzz Input Generator} generates new inputs to use for fuzz tests. The \textbf{Test Executor} converts the test inputs into actual calls to the service under test, which is invoked through the Binder RPC interface of the service.

It is important to note that the input surface tested by this solution is invokable by any process and application of the Android system. This fact has security implications, as a malicious app could leverage a bug to crash the Android services or, even worse, the bug can be the basis for gaining privileges and for obtaining confidential data (\cfr \S~\ref{subsec:bugs}). Since malicious apps are a real, critical security threat for mobile devices, it becomes important to perform testing with invalid inputs to Android services as in this work. Even if we do not consider security concerns, testing with invalid inputs is still important to prevent accidental failures that could occur in practice. Android services are the basis for a variety of mobile apps, including not only stock apps from the vendor, but also apps of uncertain quality from third-party developers. Since the interface of Android services is exposed to many service consumers not known in advance, there are many opportunities for bad data values to be circulated within the system, thus increasing the importance of gracefully handling such exceptional inputs.

\subsection{Extended architecture of the \chizpurfle{} platform}
\label{subsec:extended_arch}

\begin{figure*}
  \centering
  \includegraphics[width=\textwidth]{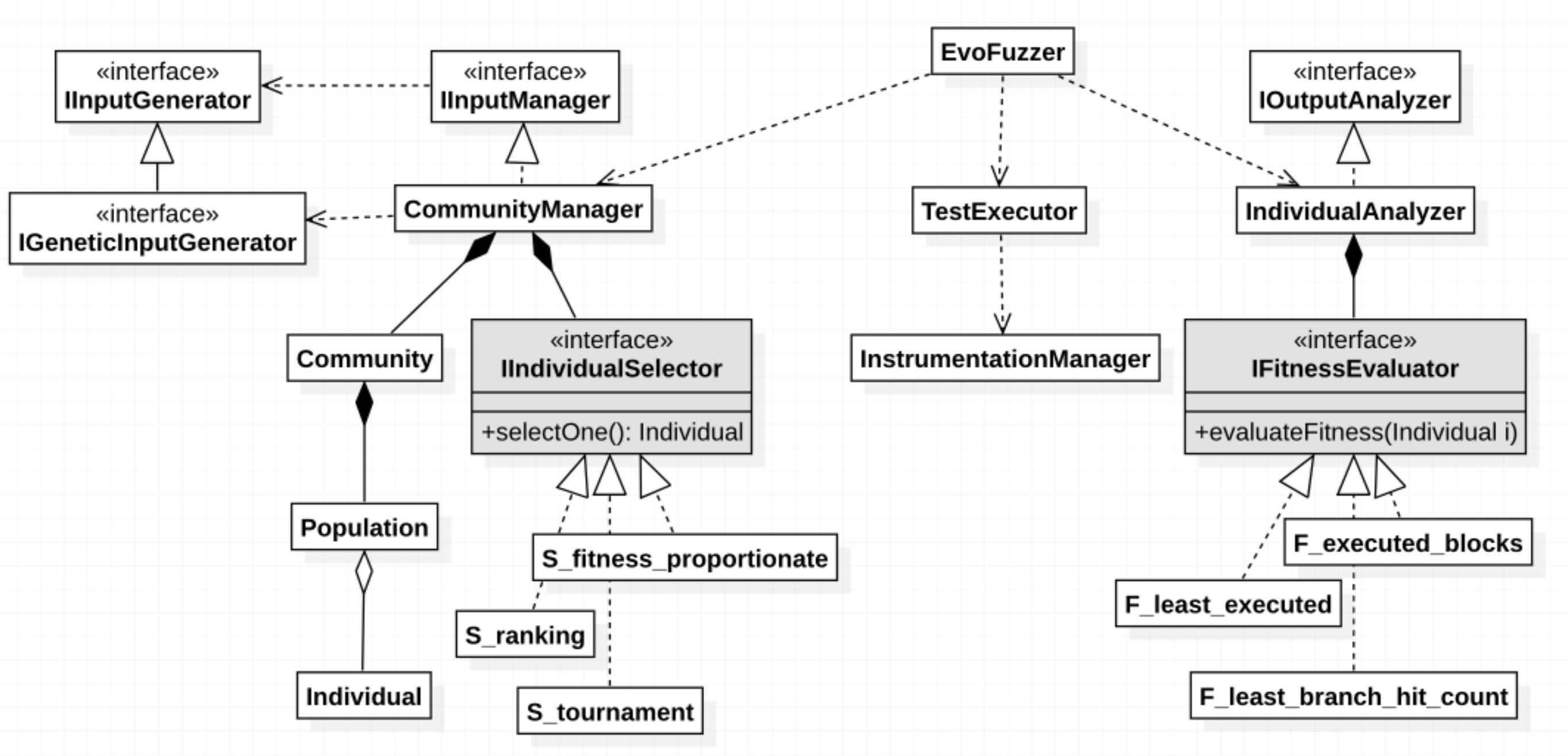}
  \caption{Architecture of the Chizpurfle platform.}
  \label{fig:evo_platform_class}
\end{figure*}

The \chizpurfle~platform has been extended in this work to support the inclusion of different evolutionary algorithms for test generation. \figurename{}~\ref{fig:evo_platform_class} shows the architecture of the platform. The grayed-out interfaces enable the implementation of new fitness functions and selection algorithms that can be plugged in the \chizpurfle~platform.
The \chizpurfle~platform includes explicit representations for individuals, populations, and communities. The Input Generator, that in the previous version only supported random and mutation operators, has been extended to also support the crossover operator, which is peculiar of genetic algorithms. The Community Manager, which implements the Input Manager interface, uses these generators and requires an Individual Selector. The Individual Selector must implement a selection algorithm necessary to choose  individuals that will be used as parents for the next generation. 
The Output Analyzer is implemented by an Individual Analyzer, which uses a Fitness Evaluator to assign fitness values to individuals based on coverage information.

\subsection{Evolutionary Fuzzing Algorithm}
\label{subsec:evo_fuzz_alg}

The general evolutionary fuzzing approach of \chizpurfle{} is presented in Algorithm~\ref{alg:chizpurfle_evo_main}.

\begin{algorithm}
\caption{Evolutionary fuzzing}
\label{alg:chizpurfle_evo_main}
\begin{algorithmic}[1]
\renewcommand{\algorithmicrequire}{\textbf{Input:}}
\renewcommand{\algorithmicensure}{\textbf{Output:}}
\REQUIRE Service s, Condition stopCondition
\STATE instrumentProcess(s) \label{lst:line:instrument}
\STATE community = new Community(s) \label{lst:line:new_community}
\FORALL{population \textbf{in} community} \label{lst:line:create_ind}
\STATE population.createRandomIndividuals()
\STATE population.offspring=\{\}
\ENDFOR \label{lst:line:create_ind:end}
\WHILE{\textbf{not} stopCondition} \label{lst:line:condition}
\FORALL{individual \textbf{in} community} \label{lst:line:forall}
    \STATE executeTest(individual) 
    \STATE analyzeTest(individual) 
    \STATE saveTest(individual) 
    \ENDFOR \label{lst:line:forall:end}
\STATE community.updateFitnessValues() \label{lst:line:update_fitness}
\STATE community.updateTargetSizes() \label{lst:line:update_targetsize}
\FORALL{population \textbf{in} community} \label{lst:line:generation}
    \REPEAT
      \STATE individual = population.selectOne()\label{lst:line:select1}
      \IF{randomProbability() < crossoverRate} \label{lst:line:crossover}
            \STATE ind2 = population.selectOne()\label{lst:line:select2}
          \STATE individual = crossOver(individual, ind2)
          \ENDIF \label{lst:line:crossover:end}
      \IF{randomProbability() < mutationRate} \label{lst:line:mutations}
              \STATE individual = mutate(individual)
      \ENDIF \label{lst:line:mutations:end}
      \STATE population.addOffSpring(individual)
    \UNTIL{(population.OffSprings.size ==\\\hspace*{1cm}population.targetSize)}
    \STATE population.individuals=population.offsprings\label{lst:line:new_population}
    \STATE population.offsprings=\{\}\label{lst:line:new_population:end}
    \ENDFOR \label{lst:line:generation:end}
\ENDWHILE 

\end{algorithmic}
\end{algorithm}

In line~\ref{lst:line:instrument}, \chizpurfle~looks for the Android OS process that hosts the service under test, and instruments the process to gather runtime information during the tests. Then, in line~\ref{lst:line:new_community}, it creates the community for the service under test, by creating a population for each public method in the service interface.

Lines~\ref{lst:line:create_ind}\mytilde\ref{lst:line:create_ind:end} initialize the populations with random individuals, as in \emph{generation-based} fuzzing tools. The number of individuals in each population is a parameter of the evolutionary algorithm, namely \code{populationInitialTargetSize}. The individuals are randomly generated according to the type of the arguments for the method. 
Alternatively, it is also possible to start from inputs recorded during the normal execution of the device, or from existing test cases. However, since we focus on proprietary Android OS services, we do not have test cases from the vendor developers. Moreover, it is difficult, and still not sufficient, to use initial inputs captured when running the services, since proprietary customizations may be exercised by specific apps or UI events that do not occur when running generic workloads for the Android OS.

\chizpurfle~evolves the community until a \code{stopCondition} is met (line~\ref{lst:line:condition}). This stop condition can be a time condition (\eg stop after 2 hours of testing), a generation limit (\eg stop after the 10th generation), or a specific objective (\eg stop after 3 failures).

The evolutionary loop starts by executing, analyzing, and saving the results for every individual in the community (lines~\ref{lst:line:forall}\mytilde\ref{lst:line:forall:end}). During the execution, the Instrumentation Module of \chizpurfle~saves a list of covered blocks and branches, which are analyzed together with the Android logs and any exception raised by the method call. This information is also saved on the device storage for off-line analysis.

Line~\ref{lst:line:update_fitness} computes the fitness function for each analyzed individual. The \keyword{fitness function} is a heuristic that ranks individuals with respect to their potential for generating (through mutation and cross-over) good new individuals that can cover more code. We included in \chizpurfle~the following three fitness functions from previous fuzz testing studies, where the fitness value is proportional to:

\begin{itemize}

\item \emph{The number of executed blocks} (\code{$F_{executed\_blocks}$}): the fittest individual is the input that executed more blocks, as evolving an individual that exercises long paths has the potential to trigger subtle bugs;

\item \emph{The number of least executed blocks} (\code{$F_{least\_executed}$}): the fittest individual is the input that executed those blocks that have been executed the least by the previous generations, in order to promote the exploration of new paths \citep{bohme2016coverage};

\item \emph{The number of times the executed branches have been executed by other individuals} (\code{$F_{least\_branch\_hit\_count}$}): the fittest individual is the input that executed those branches with the lowest  branch hit count \citep{afl2016}; the hit count is tracked with logarithmic bins, as in the AFL fuzzing tool (\ie the number of inputs that executed the branch n times, with $2^k \leq n < 2^k -1$).

\end{itemize}

Once every individual has a fitness value, \chizpurfle~computes the new target size for each population, \ie how many individuals should the next iteration generate for the population  (line~\ref{lst:line:update_targetsize}). Our goal is to boost the testing of promising methods and to avoid wasting time on trivial methods. To this goal, we reward the population with a higher average fitness value, by increasing its size by 1, and by reducing by 1 the weakest population (with a minimum of two individuals). 
Moreover, \chizpurfle~includes a second mechanism that modifies the target sizes, which serves to avoid populous communities to take a very long time for an iteration. When the total number of individuals exceeds the \code{maxCommunitySize} parameter by a certain amount (a surplus), \chizpurfle~reduces the target size of the worst populations by half of the surplus to make the total number of individuals converge at \code{maxCommunitySize}.

The next step is the generation of new individuals (\ie offsprings), to reach the target size of each population (lines~\ref{lst:line:generation}\mytilde\ref{lst:line:generation:end}). 
For each offspring, an individual is first selected (line~\ref{lst:line:select1}). With a probability \code{crossOverRate}, the individual is mixed with a second one with a \keyword{crossover} operation (lines~\ref{lst:line:crossover}\mytilde\ref{lst:line:crossover:end}). \chizpurfle~randomly applies one of these crossover operators:

\begin{itemize}

\item \emph{single-point}: A single crossover point on both parents is chosen. The offspring takes the part before the point from the first parent, and merges it with the part beyond the point of the second parent;

\item \emph{two-points}: Two points are chosen on the parents. The offspring takes the inner part from the first parent, and the outer part from the second parent;

\item \emph{uniform}: the offspring randomly takes every single point from one parent or the other.

\end{itemize}

Crossover operators are applied in a cascade mode. First, the operators select a point among the parameters of the method, where every parameter represents a point. Afterwards, the crossover operator focuses on the selected parameter, by selecting a point inside the value of the parameter. In this case, the nature of the point depends on the parameter type: For example, for primitive types, we consider their binary representation, where every bit represents a point; in the case of strings, every character represents a point; and, for complex objects, every public field represents a point. 

In a similar way, with a probability \code{mutationRate}, the offspring is mutated (lines~\ref{lst:line:mutations}\mytilde\ref{lst:line:mutations:end}). The fuzz operators include:

\begin{itemize}
\item \emph{Primitive types} (boolean, byte, char, double, float, integer, long, short): substitute with a random value, substitute with the additive identity (0), substitute with the multiplicative identity (1), substitute with the maximum value, substitute with the minimum value, add a random delta, subtract a random delta, substitute with a special character (only for char);
\item \emph{Strings}: substitute with a random string, substitute with a very long random string, truncate the string, add random substring, remove random substring, substitute a random character from the string with a special character, substitute with the empty string, substitute with null;
\item \emph{Arrays} and \emph{Lists}: substitute with an array of random length and items, remove random items, add random items, apply fuzz operator on an item value according to its type, substitute with an empty array, substitute with null; 
\item \emph{Objects}: substitute with null, invoke constructor with random parameters, apply fuzz operator on a field value according to its type.
\end{itemize}

For a given parameter type, one of the listed operators is randomly chosen each time.
For \emph{Object} types, ad-hoc fuzzers exist for important specific classes defined by the Android OS. For example, the \emph{android.content.Intent} class has a specific fuzzer that injects into the fields of an Intent (such as actions, categories, and extras) special values that have a meaning for the Intent (\emph{e.g.,} ACTION\_MAIN and ACTION\_CALL for the Intent actions) \citep{androidintent}; and the fuzzer for the \emph{android.content.ComponentName} class takes into account which components are installed on the target device, in order to use and to mutate valid component names during fuzz testing. For all the other classes, a generic object fuzzer uses the Java Reflection API to create new objects using the class constructor with random parameters, and to invoke \emph{setter} methods of the class to place random values in the fields of the object.

Every time an individual is selected in the generation phase (lines~\ref{lst:line:select1} and \ref{lst:line:select2}), a \keyword{selection algorithm} is in charge of this choice. \chizpurfle~implements three algorithms: 

\begin{itemize}

\item \code{$S_{fitness\_proportionate}$}: individuals are chosen with a probability proportional to their fitness value (inspired by B{\"a}ck \etal\citep{back1996evolutionary});

\item \code{$S_{ranking}$}: individual are chosen according to its position in the rank of individuals (inspired by B{\"a}ck \etal\citep{back1991extended});

\item \code{$S_{tournament}$}: a number \code{tour} of individuals is chosen randomly from the population, and the best individual from this group is selected as a parent, for each requested parent (inspired by Goldberg \etal\citep{goldberg1991comparative}). 

\end{itemize}

Finally, in lines~\ref{lst:line:new_population}\mytilde\ref{lst:line:new_population:end}, the offsprings take the place of the individuals of the previous generation, and are used for the next iteration of the algorithm.

\section{Experimental analysis}
\label{sec:results}

This section presents experiments with the proposed \chizpurfle~platform, according to the research questions that were introduced in Section~\ref{sec:rq}. We first compare the evolutionary approach with the simpler black-box approach (Subsections~\ref{subsec:evo_bb} and \ref{subsec:bugs}). Then, we evaluate the different options for the fitness function and the selection algorithm (Subsection~\ref{subsec:evo_configs}).

\begin{table*}[!ht]
\centering
\caption{Descriptive statistics of the tested Android system services}
\label{tab:services}
\begin{tabular}{ccc@{\hskip 15pt}cccccc}
\toprule
	&	\multirow{3}{*}{\textbf{Service}}	&	\multirow{3}{*}{\textbf{\# Methods}}	&	\multicolumn{3}{c}{\textbf{\# All parameters}}	&	\multicolumn{3}{c}{\textbf{\# Object parameters}} \\
\cmidrule{4-9}
					 &  			&              & \emph{Mean}            & \emph{Min}   & \emph{Max}   & \emph{Mean}               & \emph{Min}     & \emph{Max}    \\\midrule
\multirow{5}{*}{\rotatebox[origin=c]{90}{\parbox{1.6cm}{\centering  HUAWEI}}}  & BastetService    & 28      & 2.54   & 1 & 8   & 0.71     & 0     & 5   \\
                         & hwAlarmService   & 2      & 4          & 3   & 5   & 1               & 1    & 1   \\
                         & hwConnectivityExService & 2      & 1            & 1   & 1   & 0.5                & 0    & 1   \\
                         & hwUsbExService    & 1       & 2             & 2   & 2   & 1              & 1     & 1    \\
                         & nonhardaccelpkgs  & 4     & 1.25            & 1   & 2   & 1                & 1   & 1  \\\midrule
\multirow{5}{*}{\rotatebox[origin=c]{90}{\parbox{1.6cm}{\centering  LG NEXUS}}}   & ethernet      & 3      & 1           & 1   & 1   & 1            & 1     & 1   \\
                         & ims       & 8       & 2            & 1   & 4  & 0.5                & 0    & 2    \\
                         & isms       & 20      & 4.45            & 1  & 8   & 2.3                & 0     & 6    \\
                         & phone      & 68    & 1.5             & 1.0   & 5   & 0.72     & 0     & 4    \\
                         & sip        & 6       & 2.5             & 2   & 4  & 2.5                & 2     & 4   \\\midrule
\multirow{5}{*}{\rotatebox[origin=c]{90}{\parbox{1.6cm}{\centering   SAMSUNG}}} & ABTPersistenceService & 14      & 3.14   & 1   & 8   & 2.5                & 0     & 6    \\
                         & CocktailBarService   & 41      & 1.8   & 1   & 4   & 0.88     & 0     & 3    \\
                         & knoxcustom      & 142    & 1.41   & 1   & 5   & 0.45     & 0     & 4    \\
                         & spengestureservice  & 6       & 2            & 1  & 5   & 0.83     & 0     & 2    \\
                         & wifihs20      & 9       & 1.11   & 1  & 2   & 0.67     & 0     & 1  \\\bottomrule

\end{tabular}
\end{table*}

The tests were performed on three high-end Android smartphones, \ie a \textbf{Huawei P8} (running Android 6), an \textbf{LG Nexus 5X} (running Android 7), and a \textbf{Samsung Galaxy S6 Edge} (running Android 7). For each device, we use the \chizpurfle{} Method Extractor to detect system services customized by the vendor. 
The \chizpurfle{} Orchestrator ran on a MacBook Pro with 3GHz Intel Core i7 processor and 8 GB 1600 MHz DDR3 memory. The other components of \chizpurfle{} ran on the actual Android devices, in order to be able to test their proprietary customizations.

Overall, the three devices expose 45 custom services in Huawei, 134 in Samsung, and 10 in LG Nexus. The Chizpurfle platform leverages the Java Reflection API to get information about the methods and parameters of Android services. Therefore, we focus on the Android services that are implemented in Java, which represent a large majority of services: 32 custom services in Huawei, 114 in Samsung, 5 in LG Nexus. We performed tests on 5 Android services per device (15 target services in total), in order to have a uniform number of target services across the devices, and for the following two reasons.

First, every fuzz testing campaign has a large computational cost, as it consists of a large number of tests (where a \emph{test} means to generate a set of test input parameters, perform one method invocation using these parameter inputs, gather coverage, and assess the occurrence of failures). 
In our context, the computational bottleneck is represented by the Android device in which we run the tests: this is an unavoidable limitation, since our approach is meant to test proprietary services running on the actual Android device (e.g., we cannot simply run the proprietary version of the Android OS on a device emulator on a more powerful machine). 
Therefore, performing the analysis on all of the 151 proprietary Android services (i.e., ten times the services that we tested in this paper) would incur in an unaffordable computational cost. Focusing on a subset of services led to more than 8 millions of experiments in 65 days 4 hours 19 minutes (CPU time).

Second, we selected those services with a testing surface quite diverse from each other, both in terms of number of methods, and in terms of type and number of parameters to the methods. In order to get the right services, we selected the set of services randomly, and then checked that the services complied with our requirement of having diverse targets.  
After the tests, we found that this diversity was able to point out both the best and the worst cases for the proposed approach (\eg cases with a tenfold increase of coverage, and cases with a coverage equivalent to black-box fuzzing), and allowed us to discuss the strengths and the limitations of the proposed approach in this section.

We report descriptive statistics for the complexity of the system services in \tablename~\ref{tab:services}, including the number of the public methods of the service, the number of arguments, and the number of objects (i.e., non-primitive types) among the arguments. 
Some Android services tend to expose few methods, but their input parameters can be complex objects. For example, the ``\emph{spengesture}'' service (also discussed in Subsection~\ref{subsec:bugs}) can take in input an array of objects representing UI events; and the ``\emph{sip}'' service takes in input a SIP URI string whose format is complex and prone to string parsing vulnerabilities. In other cases, the services can expose tens, or even hundreds of methods (such as 142 methods in the case of ``\emph{knoxcustom}''). We consider a combination of services with different amounts and complexity of methods, and we discuss how the complexity of the services relates to evolutionary fuzzing.

It is important to note that the total number of basic blocks in each target service is not available to us, since the Android OS architecture does not clearly separate the code of distinct services into distinct processes and binary executables. Instead, in the Android OS architecture, the services (including proprietary ones) run as threads within the same shared process. For example, the \emph{Media Server} process of the Android OS runs both the \emph{Camera Service} and the \emph{Audio Flinger} services; the System Server runs a large number of privileged services (several tens of services in the AOSP \citep{stackoverflow_systemserver, yaghmour2013embedded}), such as the Activity Manager, the Package Manager, the Power Manager, and many others. In addition to AOSP services, Android vendors introduce their own proprietary services (and threads) to these processes. 
Since the thread of the target service shares (and, in principle, can execute) the whole code area of the process where they run with all other threads in the process (\ie the code of all services that were compiled into the binary executable of the process), it is difficult to know which specific subset of the code area belongs to a specific target service. This problem is exacerbated by the lack of the source code for the proprietary services (\ie we cannot simply infer the code belonging to a service by looking at how the source code is organized in different source code files/folders). For these reasons, we cannot accurately know which are the basic blocks of code that can be potentially reached by a target service. This is one of the reasons that makes proprietary services difficult to fuzz, since it is difficult to detect whether the test coverage is saturating the code of a target service.

We remark that even if we considered several devices from different vendors, our purpose is not to compare them, but to evaluate the effectiveness of fuzzing techniques in different contexts. Moreover, most of the proprietary services are only available on specific devices and do not have equivalent counterparts among the competitors.

\subsection{Test coverage achieved by evolutionary and black-box fuzzing}
\label{subsec:evo_bb}

We first tested each service with the evolutionary approach under the simplest configuration. We adopted the \emph{number of executed blocks} (\code{$F_{executed\_block}$}) as fitness function, and the \emph{fitness proportionate reproduction} (\code{$S_{fitness\_proportionate}$}) as selection algorithm. We tuned the \chizpurfle{}'s parameters (as shown in \tablename~\ref{tab:evo_parameters}) based on best practices and past experiences in the fields of fuzz testing and genetic algorithms  (\cfr~Section~\ref{sec:related_work}).

Then, we tested the same services with black-box fuzzing, by configuring the \chizpurfle{} platform to only use mutation operators (\ie \code{crossOverRate} drops to 0\% and \code{mutationRate} becomes 100\%) and to disable the coverage-guided feedback loop. Therefore, individuals are randomly mutated, populations do not change in size, and no method is preferred over the others. 
Since input generation is not driven by test coverage, we do not perform dynamic binary instrumentation, and disable the Instrumentation Module. 
To perform a fair comparison between the evolutionary and black-box approaches, we ran black-box fuzzing for the same amount of time of evolutionary fuzzing. 
However, the black-box approach does not provide us coverage data for performing a comparison. Therefore, after the execution of black-box fuzzing, we replayed the same black-box tests for a second time, and used the dynamic binary instrumentation technique to collect coverage data for the black-box tests. In this second run, we enabled the Instrumentation Module during black-box fuzzing, even if this component is not applied for driving input generation. 
By running the black-box tests respectively without and with instrumentation, we can first generate the same amount of tests that would be produced by a non-coverage-driven black-box approach, and then evaluate the coverage for these tests in a separate phase.

Even if dynamic binary instrumentation introduces a slow-down on the speed of test execution, the slow-down still allows the Android services to execute without any noticeable side effect, thus preserving the intended behavior of the tests.
On average, the slow-down of coverage profiling is 13.91x. 
To put this number into context, we must consider that the performance slow-down is inline with other tools for dynamic binary instrumentation. For example the Valgrind framework (which also uses dynamic binary rewriting for complex analyses, such as finding memory leaks and race conditions), when applied on the SPEC CPU 2006 benchmark \citep{nethercote2007valgrind}, causes an average slow-down of 4.3x when the program is simply executed on the Valgrind virtual machine; and an average slow-down of 22.1x when performing memory leak analysis. Such overhead when running tests is rewarded by a higher bug-finding power, and it is in many cases accepted by developers as shown by the widespread adoption of Valgrind in automated regression test suites in open-source projects \citep{cotroneo2013fault}. 
The comparison between black-box and evolutionary fuzzing aims to to evaluate whether the slow-down induced by coverage profiling is rewarded by a more thorough coverage.

\begin{table}[!ht]
\centering
\caption{Parameters for the genetic algorithm of Chizpurfle}
\label{tab:evo_parameters}
\begin{tabular}{cc}
\toprule
\textbf{Parameter} & \textbf{Value} \\\midrule
\code{populationInitialTargetSize} & 10 \\
\code{stopCondition} & 20th generation\\
\code{maxCommunitySize} & 200\\
\code{crossOverRate} & 80\% \\
\code{mutationRate} & 5\% \\
\code{tour} & 5 \\\bottomrule 
\end{tabular}
\end{table}

We executed 10 repetitions of both the evolutionary and black-box fuzzing campaigns in order to gain statistical confidence in the results, in a similar way to previous studies in the field of random testing \citep{arcuri2014hitchhiker}. 
We performed a one-way analysis of variance (ANOVA) to analyze how the choice between black-box and evolutionary fuzzing affects the coverage of the service under test. We also used the service name as a block-factor, \ie a second factor, which forms a two-way analysis without interaction. 

We consider the number of basic blocks as the dependent variable. The one-way ANOVA tests the null hypothesis that there are no differences between the mean values of two or more independent groups. If it returns a statistically significant result, we reject the null hypothesis: there are at least two group means that have a statistically-significant difference.

Since the normality assumptions are not met by the dataset, we adopted the non-parametric Mann-Whitney test with the two-sample normal approximation \citep{mann1947test}, instead of the conventional Fisher test \citep{fisher1922interpretation}. The p-value, related to the null hypothesis that there is no difference between black-box and evolutionary approach, is less than 0.0001, \ie we can strongly reject the null hypothesis, and confirm that there is a difference. 
The results of the one-way ANOVA are summarized in \tablename{}~\ref{tab:anova_bbevo}.

\begin{table}[!ht]
\centering
\caption{One-way ANOVA for the evolutionary (EVO) and black-box (BB) approaches}
\label{tab:anova_bbevo}
\begin{tabular}{cc}
\toprule
groups & EVO and BB\\
number of samples per group & 150\\
\midrule
mean of EVO group& 159.4\\
std. deviation of EVO group& 263.157\\
\midrule
mean of BB group& 53.147\\
std. deviation of BB group& 133.89\\
\midrule
Mann-Whitney p-value & <0.0001\\
$\hat{A}_{EVO,BB}$ & 0.74\\
\bottomrule
\end{tabular}
\end{table}

To quantify the strength of this result, we computed the effect size for the Mann-Whitney test.
We use the Vargha and Delaney's $\hat{A}$ statistic, a non-parametric effect size measure, to compute the common language effect \citep{vargha2000critique}. The common language effect is a measure from two groups, which represents the probability of confirming a hypothesis when comparing a random pair of samples from the two groups. 
In our case, we consider the hypothesis that the evolutionary approach covers more code than the black-box approach. The Vargha and Delaney's statistic showed better results for the evolutionary approach in 74\% of the cases ($\hat{A}_{EVO,BB}=0.74$). Thus, in our experiments, in more than 7 times out of 10, the evolutionary approach covers more code than the black-box approach in the service under test.

For all services, the evolutionary approach performed the same or better than the simpler black-box approach. \tablename{}~\ref{tab:evobb_ratio} shows the ratio between block coverage of the genetic algorithm approach and black-box approach, by averaging the results of the repetitions. This result confirms that the evolutionary approach should be preferred over black-box, despite the slow-down caused by coverage profiling. However, coverage profiling may not provide a significant benefit for some services.

\begin{table}[!ht]
  \centering
  \caption{Average code coverage gain from black-box to evolutionary fuzzing (EVO-to-BB ratio)}
  \label{tab:evobb_ratio}
\begin{tabular}{@{}lll@{}}
\toprule
\textbf{Device}  & \textbf{Service}          & \textbf{Gain}      \\ \midrule
\multirow{5}{*}{HUAWEI}  & BastetService           & 1.43 \\
                         & hwAlarmService          & 2.29 \\
                         & hwConnectivityExService & 2.43  \\
                         & hwUsbExService          & 1.14 \\
                         & nonhardaccelpkgs        & 1           \\ \midrule
\multirow{5}{*}{LG}  & ethernet                & 1           \\
                         & ims                     & 1           \\
                         & isms                    & 1.97 \\
                         & phone                   & 24.28 \\
                         & sip                     & 1.14 \\ \midrule
\multirow{5}{*}{SAMSUNG} & ABTPersistenceService   & 1           \\ 
                         & CocktailBarService      & 1.01 \\
                         & knoxcustom              & 1           \\
                         & spengestureservice      & 2.14 \\
                         & wifihs20                & 2.6 \\ \bottomrule
\end{tabular}
\end{table}

The main observation is that the performance of the evolutionary approach can be related to the complexity of the service under test. The services with the lowest differences between black-box and evolutionary fuzzing are also the ones with the lowest number of methods. However, the number of methods seems not to be the only factor that is related to the highest gains from the evolutionary approach. Indeed, the complexity of the methods' signature, in terms of number of arguments, seems to affect the results. For example, the \code{isms} and \code{phone} services of the LG Nexus 5X have up to 8 and 5 arguments in their methods, resulting in a high EVO/BB ratio (\ie the ratio between the code coverage of the evolutionary approach and of the black-box approach). The \code{hwAlarmService}, whose methods have up to 5 arguments, exhibits a high ratio despite the low number of methods.

The \code{ABTPersistenceService} of the Samsung Galaxy S6 Edge is the main exception to this interpretation, since it has a large number of methods with up to 8 arguments, but a low EVO/BB ratio. We believe this is due to the high mean in the number of objects among its arguments (\ie 2.5), since this low ratio may have been caused by the simplistic approach adopted by \chizpurfle{} to fuzz non-standard objects; therefore, researching new, more complex fuzzing operators represents another interesting research direction. 

Finally, we should also consider the nature of the service. For example, the Nexus \code{Ethernet} service is probably used to manage wired Ethernet connections, which is enabled only with a special adapter physically connected to the device. Thus, most of this service code is hard to reach with both approaches, resulting in an EVO/BB ratio equal to 1.

\begin{table}[!ht]
  \centering
  \caption{Average code coverage gain from black-box to evolutionary fuzzing (EVO-to-BB ratio), without the community mechanism}
  \label{tab:evobb_ratio_nocommunity}
\begin{tabular}{@{}lll@{}}
\toprule
\textbf{Device}  & \textbf{Service}          & \textbf{Gain (no community)}      \\ \midrule
\multirow{5}{*}{HUAWEI}  & BastetService           & 1.23 \\
                         & hwAlarmService          & 2.29 \\
                         & hwConnectivityExService & 2.43  \\
                         & hwUsbExService          & 1.14 \\
                         & nonhardaccelpkgs        & 1           \\ \midrule
\multirow{5}{*}{LG}  & ethernet                & 1           \\
                         & ims                     & 1           \\
                         & isms                    & 1.01 \\
                         & phone                   & 12.97 \\
                         & sip                     & 1 \\ \midrule
\multirow{5}{*}{SAMSUNG} & ABTPersistenceService   & 1 \\ 
                         & CocktailBarService      & 1.01 \\
                         & knoxcustom              & 1           \\
                         & spengestureservice      & 1.82 \\
                         & wifihs20                & 2.13 \\ \bottomrule
\end{tabular}
\end{table}

\begin{figure}[!ht]
  \centering
  \includegraphics[width=\textwidth]{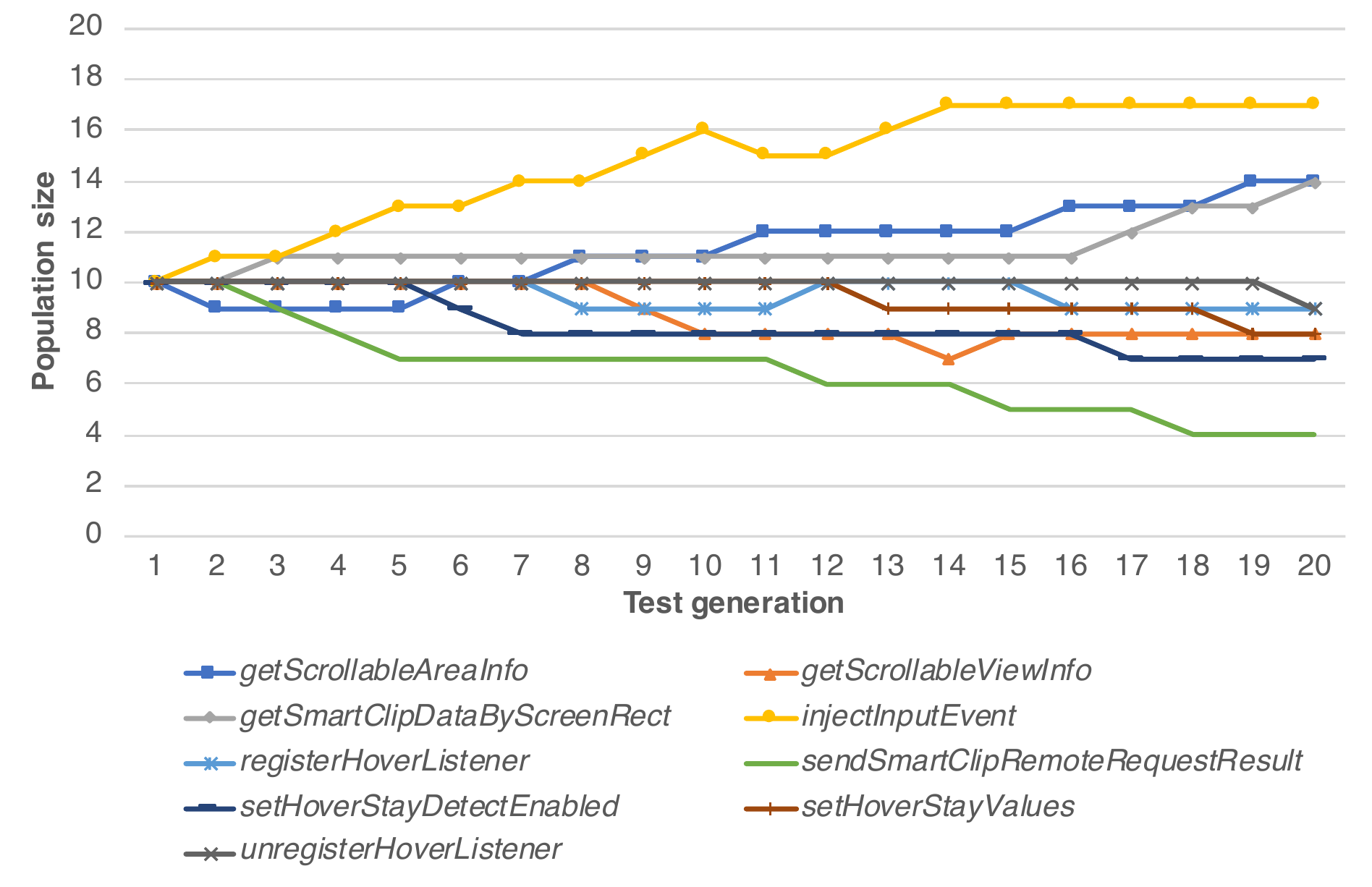}
  \caption{Evolution of population size in the community for the \emph{spengesture} service.}
  \label{fig:example_community}
\end{figure}

To further analyze the impact of the idea of \emph{community} to co-evolve test inputs, we performed additional tests for comparing black-box and evolutionary fuzzing. In these tests, we used the same configuration of the previous round of experiments, but we disabled the \emph{community} mechanism. Therefore, the size of the populations of test inputs has been equal across all methods, and fixed throughout the duration of the fuzz testing campaigns.

We report in \tablename{}~\ref{tab:evobb_ratio_nocommunity} the gain in coverage between black-box and evolutionary fuzzing without the \emph{community} mechanism. For some services, the gain was the same of the case of evolutionary fuzzing with the \emph{community} mechanism (i.e., the services exhibit the same gains in \tablename{}~\ref{tab:evobb_ratio} and \ref{tab:evobb_ratio_nocommunity}), whereas for other services there was a reduction of the gain due to the lack of the community mechanism (i.e., the gain in \tablename{}~\ref{tab:evobb_ratio_nocommunity} was lower than \tablename{}~\ref{tab:evobb_ratio}). We remark that the \emph{community} mechanism brings an improvement (up to doubling the coverage gain for the \emph{phone} service) for the services that expose a large number of methods (as reported in \tablename{}~\ref{tab:services}). These larger services have more methods whose coverage is quickly saturated, thus giving to the \emph{community} mechanism more opportunities to reduce the population of such methods in favor of more promising ones. \figurename{}~\ref{fig:example_community} shows an example of evolution of the population sizes for one of the services under test (\emph{spengesture}), where the population size of many methods is reduced in favor of the \emph{injectInputEvent} method.


\subsection{Robustness bugs found by evolutionary fuzzing}
\label{subsec:bugs}

During evolutionary fuzzing, the \chizpurfle{} platform incurred in critical failures of privileged OS services, thus with a strong impact on the reliability of the Android device. Moreover, these failures can be potentially leveraged to get confidential information and to subvert key Android processes; therefore, they have a high value for malicious attackers. These failures were triggered by complex combinations of inputs, which were generated after few generations of fuzzing tests. We note that the community mechanism was unable to further increase the time to find these vulnerabilities, since it takes some generations to gradually develop during the fuzz campaign (e.g., in \figurename{}~\ref{fig:example_community}, the population sizes varies over a period of more than 10 generations).

To analyze the failures, we first ensured that the failures were reproducible, by re-issuing the inputs that triggered the failures during fuzzing, and checking that the failures occur again. Then, we analyzed the input surface of the vulnerable target service, and the failure messages that the service reported on the logs, which included uncaught exceptions and the stack trace at the time of the failure. Despite the source code was not available to us, this information allowed us to understand the root cause of the failures, and trace the failures back to 2 distinct bugs.

The first bug showed up while testing the \emph{spengestureservice} service of the Samsung device, hosted by the \emph{system\_server} process. The individuals that made the tests fail belong to the \emph{injectInputEvent} method population. Even if we cannot access to the source code of this method, we found a similar method (with the same name and minor differences in the method signature) provided by the \emph{InputManager} class of AOSP, which handles input devices such as keyboards. This method ``injects an input event into the event system on behalf of an application'' \citep{injectinputevent}. It is likely that the method with the same name in the \emph{spengestureservice} performs the same operation for input events from the ``S Pen'' in Samsung devices \citep{spen}. The method gets as parameter, among others, an array of \emph{android.view.InputEvent} objects, which is an abstract class for representing input events from hardware components. \chizpurfle{} detected a \emph{FATAL EXCEPTION} when this array is non-null and non-empty, and at least one of its elements is null (instead, the service does not fail if the array is simply null or empty). This input causes the service to throw a \emph{NullPointerException} that is not caught, causing a crash. Depending on which process consumes the injected events, the bug can manifest as either a crash of the \emph{com.android.systemui} process with a black screen of the user interface for a few seconds; or a crash of the \emph{system\_server} process with the restart of the whole Android device.

The second bug belongs to the \emph{callInVoIP} method from the \emph{voip} service of the Samsung device. Very likely, this method is used by the VoIP app for corporate users from Samsung \citep{wevoip}. The test input is a SIP address URI, represented as a long, structured string (\eg ``sip:1-999-123-4567@voip-provider.example.net''). If SQL control expressions are injected in this string (\eg single quotes as in SQL injection), the string triggers an \emph{SQLLiteException} that is not caught by the host process, namely \emph{com.samsung.android.incallui}. This is a customization introduced by Samsung to the AOSP \emph{com.android.incallui}, which provides an UI handler for the activity that appears during a call, and that provides on-screen functions for handling a VoIP call. Since the exception is not caught, the \emph{com.samsung.android.incallui} process crashes and cuts off any ongoing call. Moreover, the bug could be potentially used maliciously to steal private data of the VoIP service (e.g., the list of contacts) from an unprivileged application. It is important to note that this bug could not be easily triggered with random string corruption, since the injected expressions should at specific places of the SIP address URI. While black-box fuzzing was not able to find this bug, evolutionary fuzzing was able to tune the test input generation to find the vulnerable part of the URI string.


\subsection{Comparison among evolutionary configurations}
\label{subsec:evo_configs}

We tested again the same system services, to determine whether there is a configuration in the evolutionary approach that performs better than the others. We considered every combination of the 3 fitness functions and of the 3 selection algorithms, and applied these combinations on each of the 15 system services, for a total of 135 evolutionary fuzzing campaigns. 
\tablename{}~\ref{tab:fitness_means} and \tablename{}~\ref{tab:selection_means} show the number of basic blocks covered by \chizpurfle~and grouped by, respectively, the fitness functions and the selection algorithms, counting 45 samples in each group.

\begin{table}
\centering
\caption{Number of covered basic blocks, with respect to different fitness functions}
\label{tab:fitness_means}
\begin{tabular}{ccc}
\toprule
		&	\textbf{Mean}	&	\textbf{Std. dev.} \\
\midrule
$F_{executed\_blocks}$			&	51.4889	&	32.6536	\\
$F_{least\_executed}$			&	47.8889	&	17.939	\\
$F_{least\_branch\_hit\_count}$			&	46.0889	&	23.2348	\\
\bottomrule
\end{tabular}
\end{table}

\begin{table}
\centering
\caption{Number of covered basic blocks, with respect to different selection algorithms}
\label{tab:selection_means}
\begin{tabular}{ccc}
\toprule
		&	\textbf{Mean}	&	\textbf{Std. dev.}\\
\midrule
$S_{fitness\_proportionate}$			&	52.9778	&	32.61	\\
$S_{ranking}$			&	52.3111	&	23.0925	\\
$S_{tournament}$			&	40.1778	&	15.5087	\\
\bottomrule
\end{tabular}
\end{table}

We consider the null hypotheses that the choices of the fitness function and of the selection algorithm have no effect on the coverage achieved evolutionary testing, and we performed the non-parametric Kruskal-Wallis test \citep{kruskal1952use} since data are not normally distributed\footnote{Kruskal-Wallis test extends the Mann–Whitney test for more than two groups.}. We once again used the service name as a block-factor. The p-values, computed by the Chi-Square approximation \citep{mann1947test}, are 0.9693 for the fitness function factor and 0.0037 for the selection algorithm factor. 

The selection algorithm factor has a very low p-value: the choice of a selection algorithm has a statistically significant effect. However, while the ANOVA test confirms that at least two groups have a statistically-significant difference, it does not point out which specific group is the best one. 
To detect the best selection algorithm for evolutionary fuzzing, we performed pairwise tests and measure effect size in each case. The results are in \tablename~\ref{tab:pairwise_selection}. 
While the fitness proportionate reproduction (\code{$S_{fitness\_proportionate}$}) and ranking (\code{$S_{ranking}$}) selection algorithms are equivalent in terms of coverage, they both perform better than the tournament selection algorithm (\code{$S_{tournament}$}).

\begin{table*}
\centering
\caption{Pairwise analysis of \emph{selection algorithms}}
\label{tab:pairwise_selection}
\begin{tabular}{cccc}
\toprule
\textbf{Selection alg. 1} & \textbf{Selection alg. 2} & \textbf{Mann-Whitney p-value} & $\hat{A}_{alg. 1, alg. 2}$\\\midrule
$S_{fitness\_proportionate}$ & $S_{ranking}$ & 0.9742 & 0.498\\
$S_{fitness\_proportionate}$ & $S_{tournament}$ & 0.0011 & 0.7\\
$S_{ranking}$ & $S_{tournament}$ & 0.004 & 0.681\\\bottomrule
\end{tabular}
\end{table*}

\begin{figure}
  \centering
  \includegraphics[width=3.5in]{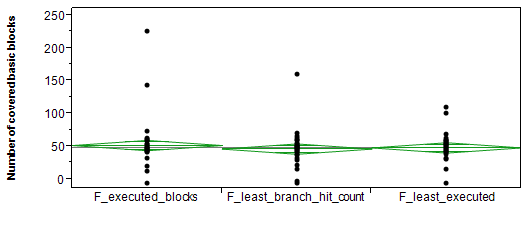}
  \caption{ANOVA visualization for the \emph{fitness function} factor}
  \label{fig:anova_fitness}
\end{figure}

\begin{figure}
  \centering
  \includegraphics[width=3.5in]{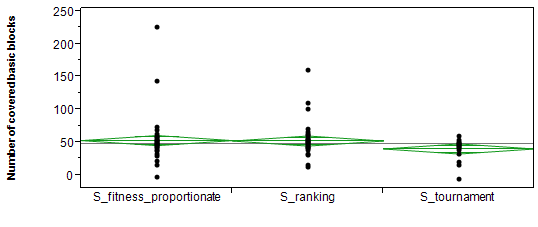}
  \caption{ANOVA visualization for the \emph{selection algorithm} factor}
  \label{fig:anova_selection}
\end{figure}

The \figurename{}~\ref{fig:anova_fitness} and \figurename{}~\ref{fig:anova_selection} show the ANOVA data and results. A means diamond represents the group mean (the line across each diamond) and the 95\% confidence interval (the vertical span of each diamond). Each diamond has also two marks above and below the group mean. For groups with equal sample sizes, as in our cases, overlapping marks indicate that the two group means are not significantly different at the 95\% confidence level. Thus, we can visually confirm that with respect to the selection algorithm factor, we must reject the null hypothesis because the tournament selection algorithm (\code{$S_{tournament}$}) diamond is right below the other two as shown in \figurename{}~\ref{fig:anova_selection}.

If we consider the combination of the fitness function and the selection algorithm as a unique factor, the non-parametric Kruskal-Wallis test returns a very low p-value of 0.0011: the Fitness Function/Selection Algorithm combination groups have a statistically significant difference. Thus, we can compute the best configuration, using the approach proposed by Fraser \etal based on configuration rank \citep{fraser2011not}. With a total of 9 configurations, we compare each configuration with the other 8. For each comparison in which a configuration is statistically better, its score is increased by one. Otherwise, if a configuration is statistically worse, its score is decreased by one. Considering an initial score of 0, the higher score a configuration has, the better it is in terms of coverage. 
The score rank is presented in \tablename~\ref{tab:score_rank}. The best combination is to use a fitness function proportional to the number of times the executed branches have been executed by other individuals (\code{$F_{least\_branch\_hit\_count}$}) with the ranking selection algorithm (\code{$S_{ranking}$}). The worst combinations are, as expected, those with the tournament selection algorithm (\code{$S_{tournament}$}), that should be avoided.

Considering the nature of the selection algorithms, we must highlight that the tournament selection algorithm (\code{$S_{tournament}$}) is the only one of the three with a tunable parameter. This algorithm selects \code{tour} individuals (5, \cfr~\tablename{}~\ref{tab:evo_parameters}) randomly from the population and returns the fittest of this group. The bound to this parameter limits the potentiality of this selection algorithm in a scenario such our one, where the populations change in size during the campaign. This consideration strengthens the statistical results, suggesting that nonparametric selection algorithms should be preferred for evolutionary fuzzing.

\begin{table*}
\centering
\caption{Ranking of fuzzing configurations}
\label{tab:score_rank}
\begin{tabular}{cccccc}
\toprule
\textbf{Rank} & \textbf{Fitness function} & \textbf{Selection algorithm} & \textbf{Score} & \textbf{Mean} & \textbf{Std. dev.}\\
\midrule
1 & $F_{least\_branch\_hit\_count}$ & $S_{ranking}$ & 6 & 58 & 31.1069\\
\midrule
2 & $F_{executed\_blocks}$ & $S_{fitness\_proportionate}$ & 5 & 69.6 & 49.4919\\
\midrule
3 & $F_{least\_executed}$ & $S_{fitness\_proportionate}$ & 3 & 49 & 12.655\\
\midrule
4 & $F_{least\_executed}$ & $S_{ranking}$ & 0 & 54.4667 & 22.0389\\
\midrule
5 & $F_{least\_branch\_hit\_count}$ & $S_{fitness\_proportionate}$ & -2 & 40.3333 & 15.7567\\
\midrule
  & $F_{executed\_blocks}$ & $S_{ranking}$ & -2 & 44.4667 & 10.8181\\
7 & $F_{executed\_blocks}$ & $S_{tournament}$ & -3 & 40.4 & 15.7671\\
  & $F_{least\_branch\_hit\_count}$ & $S_{tournament}$  & -3 & 39.9333 & 15.896\\
\midrule
9 & $F_{least\_executed}$ & $S_{tournament}$ & -4 & 40.2 & 15.9535\\\bottomrule
\end{tabular}
\end{table*}

\section{Threats to Validity}
\label{sec:threats}

\subsection{Internal Validity}

This study shows that evolutionary fuzzing can test Android OS system services more thoroughly (e.g., in terms of code coverage) compared to black-box fuzzing. Furthermore, we analyzed the impact of the choice of fitness function and selection algorithm. However, we considered only three fitness functions and three selection algorithms implemented in \chizpurfle{}. There are several potential tunings among the parameters of an evolutionary algorithm but we focused on these two, and we chose only three options for each of them. We chose them among the most popular and intuitive ones \citep{bohme2016coverage,afl2016,back1996evolutionary,back1991extended,goldberg1991comparative}, but there exist other options we did not consider. Other tunings can further improve the coverage that, with our results, seems to be already better than the simple black-box approach. The design of \chizpurfle{} is highly extensible, so that new fitness functions and selection algorithms can be easily added and used for further experimentations.

Another  threat to validity is potentially represented by the initial (random) seeds that we used for the experimentation. In principle, starting from a sample of valid seeds (i.e., input vectors that adhere to the validity constraints for the methods) can further improve the code coverage for both the black-box and grey-box approaches. 
Unfortunately, crafting input vectors for proprietary Android services (e.g., by collecting inputs from real executions under some user interactions with apps) would be problematic, and the inputs would not necessarily be of better quality than random ones. One reason is that we do not have any official documentation for proprietary services; for some of them, such as the \emph{bastet} in Huawei, we could not even find any public information about the role of the service. Therefore, we cannot know which apps exercise the service and under which conditions. We would still rely on a trial-and-error approach to define an initial set of seeds, and we would likely not be able to cover all of the several methods of the services (e.g., 28 methods in the case of \emph{bastet}). Moreover, using manually-crafted seeds would still present methodological issues, since we would have a bias due to our specific choice of the workload, and would make the study less reproducible. 
Therefore, we opted for using initial random seeds. \chizpurfle{} takes into account the data type of input parameters, in order to generate inputs that are always compliant to the signature of the methods. For all the services that we tested, this sufficed to always have many valid, structured inputs (e.g., inputs that were processed by the services without raising any exception) within the few initial generations of fuzzing (with both the black-box and gray-box approaches). The evolutionary approach further improves over the initial seeds, such that to overcome the initial checks performed by input parsers and to explore the target service in depth. Exploring more seed generation strategies is a potential direction for future research based on our fuzzing platform.

\subsection{External Validity}

This study focuses on three different Android devices, but it can be easily replicated on other Android devices. In this study, we randomly chose 5 services as target for each of the device. They represent only a fraction of all the services and can be subject to bias. Nevertheless, they are very heterogeneous services in terms of number of methods and type of arguments, as presented in \tablename~\ref{tab:services}.
Furthermore, the open-source nature of the tool should help other researchers and practitioners to perform new study and test campaigns on any Android device, even with other evolutionary approaches we did not consider.

\section{Related work}
\label{sec:related_work}

This section gives an overview of the state-of-the-art in the broad area of fuzzing, with a focus on evolutionary approaches and on the Android OS.

Fuzzing became popular in testing system software due to its ease-of-use and its applicability on libraries, network services, command-line applications, and OS kernel interfaces. A first study by \cite{miller1990empirical} tested the UNIX system utilities by injecting random inputs to their interfaces, and it found an unexpected number of crashes, leaks, and deadlocks even when the inputs look trivial. Other researchers extended this idea by creating new OS robustness testing approaches, such as BALLISTA \citep{koopman2000exception}, MAFALDA \citep{fabre1999assessment}, the DBench project \citep{kanoun2005benchmarking}, and SABRINE \citep{cotroneo2013sabrine}.

More recently, fuzzing has been enhanced with white-box solutions, such as by leveraging symbolic execution. KLEE \citep{cadar2008klee} and SAGE \citep{godefroid2008automated} are the most famous ones. KLEE is based on a virtual machine environment, which forks every time it meets a condition, exploring both paths of every branch. SAGE uses a record\&replay framework \citep{bhansali2006framework} to negate one of the logical conditions across a (concrete) path, and generates new inputs to explore different paths. These solutions use a constraint solver to find a concrete input to fulfill all of the conditions on a path. Thus, these technologies are very resource- and time-consuming.

Gray-box fuzzing is another, more lightweight approach, based on the idea of exploiting execution data to guide the generation/mutation of fuzz inputs. This category of fuzzers adopts techniques from the research area of search-based software engineering \citep{harman2012search}: hill climbing, simulated annealing, and genetic algorithms are search-based approaches that were also used for fuzzing. 
AFL \citep{afl2016} is the most widespread fuzzing tool of this kind, which has been defined by its authors as an ``instrumentation-guided genetic fuzzer''. It exploits basic evolutionary techniques to efficiently improve the quality of fuzz inputs based on coverage measurements. By mutating previous inputs, AFL tries to discover new paths. Improvements of AFL has been presented in subsequent work: AFLFast \citep{bohme2016coverage} and AFLGo \citep{bohme2017directed}. The first exploits a Markov chain model, which specifies the probability that fuzzing an input that exercises path \emph{i} generates an input that exercises path \emph{j}. The second is a directed fuzzing solution, \ie it generates inputs with the objective of reaching a given set of target program locations efficiently.

With a different objective, SlowFuzz \citep{petsios2017slowfuzz} is a framework that automatically finds inputs that trigger worst-case algorithmic behavior in the tested binary. It exploits evolutionary search techniques together with dynamic analysis. Initially, SlowFuzz randomly selects an input to execute from a given \emph{seed corpus}, which is mutated and passed as input to the program under test. During an execution, it records profiling information, such as the total count of all instructions executed. An input is scored based on its resource usage, and it is added to the mutation corpus if the input is deemed as a slow unit.

Exploring the intersection between testing and evolutionary approaches, \cite{rawat2017vuzzer} presented the VUzzer application-aware evolutionary fuzzing strategy. VUzzer combines static analysis with mutation-based evolutionary techniques to efficiently generate inputs. It exploits dynamic taint analysis, magic-byte detection, basic block weight calculation, and error-handling code detection. Instead, \cite{veggalam2016ifuzzer} proposed IFuzzer. It exploits an evolutionary fuzzing technique and targets JavaScript interpreters. IFuzzer uses the language grammar to generate valid inputs and generates new code fragments by performing genetic operations on a test suite. 

Among evolutionary approaches, it is important to mention EvoSuite \citep{evosuite,fraser2011evosuite}, which is an automatic test suite generator with assertions for classes written in Java code. It exploits evolutionary algorithms for the generation of a whole test suite, optimizing to a coverage criterion rather than individual coverage goals. A candidate solution in EvoSuite is a test suite, consisting of a variable number of individual test cases. Each of the test cases is a sequence of method calls, exercising the unit under test and setting up complex objects in order to do so. EvoMaster \citep{arcuri2018evomaster} is its spin-off: a RESTful API automated test case generator that finds faults using the HTTP return statuses as an automated oracle.

In the general research field of fuzzing, black-/gray-/white-box fuzzing techniques are all means for generating streams of unusual inputs in order to find robustness issues, with different cost-efficiency trade-offs. On one extreme, black-box fuzzing is a ``cheap'' approach to input generation, by using randomness, grammars, or lists of known problematic inputs; it tends to be the least effective approach, but it is also the easiest one to deploy (e.g., it can be applied in the earlier stages of testing) and can achieve a decent trade-off between testing efforts and bugs found (as it is lightweight and can generate a high number of inputs). On the other extreme, white-box fuzzing generates inputs by applying advanced program analysis techniques (for example, constraint solving), which is the most computationally-heavyweight approach and is the most complex approach to deploy, but it is able to find the most subtle bugs. Gray-box fuzzing (which includes evolutionary fuzzing) is in the middle of this design spectrum, as it steers the generation of inputs by leveraging feedback from the coverage of previous inputs, and it is moderately more complex to deploy than black-box fuzzing, but it is more lightweight than white-box fuzzing. As an example of complementary use of these techniques, a study from Microsoft researchers \citep{bounimova2013billions} reports a split of 66\%-33\% of bugs found respectively by black- and white-box fuzzing during the development of Microsoft’s Windows 7. 
It is also important to clarify that ``black-box fuzzing'' and ``evolutionary fuzzing'' should be considered a complement, rather than a replacement, to black-box testing and other forms of testing. In black-box testing (such as, partition-based testing), the inputs are generated by leveraging a-priori knowledge about the partitioning of the input space, the functional features, and the user requirements. Fuzzing approaches, including ours, typically do not rely on such a-priori domain knowledge.

In Android-related research, fuzzing has been extensively used to attack network and inter-process interfaces. For example, \cite{mulliner2009fuzzing} found severe vulnerabilities in the SMS protocol. Droidfuzzer \citep{ye2013droidfuzzer} targets Android activities that accept MIME data through Intents (a higher-level IPC mechanism based on Binder); \cite{sasnauskas2014intent} developed a more generic Intent fuzzer that can mutate arbitrary fields of Intent objects. \cite{mahmood2012whitebox} adopted the white-box fuzzing approach by decompiling Android apps to identify interesting inputs, and running them on Android emulator instances on the cloud. However, these tools and similar ones  \citep{maji2012empirical,au2012pscout,yang2014intentfuzzer,hu2016fuzzy,cao2015towards,feng2016bindercracker} are mostly black-box and focus on the robustness of Android apps, which expose a different attack surface than Android system services. For example, Intents are data containers that follow a specific format, which includes an action identifier, a resource URI, and other optional fields; instead, Android system services use the Binder IPC, which is a richer, RPC-oriented IPC mechanism, which cannot be fuzzed through intents.

Our evolutionary fuzzing approach differs from the state-of-the-art in several ways. First, it is implemented on top of \chizpurfle{} \citep{iannillo2017chizpurfle} and, thus, it is the first gray-box, evolutionary fuzzing approach specifically tailored for vendor customizations in the Android OS, and overcoming the limitations of this context (\eg inability to recompile the source code, and to run the services outside the device). Our approach does not target traditional fuzzing targets, such as multimedia file parsers and small command-line programs, but aims to test Android services with an API-oriented interface. Moreover, to the best of our knowledge, it is the only approach that introduces the concept of community, enabling the co-evolution of populations with individuals physically located in the same target but impossible to combine due to syntactic constraints.

\section{Conclusion}
\label{sec:conclusion}
This paper presented a novel evolutionary fuzzing approach, specifically tailored for Android OS and its closed-source customizations by Android vendors. 
We developed a tool, \chizpurfle, to provide a general platform for evolutionary fuzzing. The main feature of \emph{Chizpurfle} is its ability to work directly on the commercial device and its compiled code. 
Moreover, we tested 15 system services from 3 commercial Android devices to evaluate the effectiveness of evolutionary fuzzing and the impact of its configuration. Our evolutionary fuzzing solution opens new possibilities for applying and experimenting with new testing strategies.


%

As for the first research question (RQ1), \emph{evolutionary fuzzing always performs no worse (in terms of code coverage) than black-box fuzzing, and in most cases, evolutionary fuzzing brings a noticeable increase}. Furthermore, we observed that the performance of evolutionary fuzzing seems related to the complexity of the service under test. The services with a richer API (in terms of number of both methods and arguments) are the ones that benefit the most from the evolutionary approach.

%

To address the second research question (RQ2), we evaluated typical choices for the fitness function and for the selection algorithm. Our analysis found that \emph{there is no statistically significant difference between the fitness functions}, but we noted that \emph{choosing the tournament selection algorithm noticeably reduces the performance of evolutionary fuzzing}. We believe that the parametric nature of this algorithm is not suitable for evolutionary fuzzing, where the size of the populations needs to change over the test generations.

\begin{acknowledgements}
This research was carried out in the frame of Programme STAR, financially supported by UniNA and Compagnia di San Paolo.
\end{acknowledgements}

\bibliographystyle{spbasic}      
\bibliography{bibliography}

\end{document}